\documentclass[fleqn,usenatbib]{mnras}

\usepackage{mathptmx}
\usepackage{txfonts}

\usepackage[T1]{fontenc}
\usepackage{ae,aecompl}

\usepackage{graphicx}	
\usepackage{amssymb}	
\usepackage[export]{adjustbox}

\newcommand{\kms}{km s$^{-1}$}

\newcommand{\cmm}{cm$^{-2}$}
\newcommand{\cmmm}{cm$^{-3}$}
\newcommand{\Lya}{Ly$\alpha$}

\newcommand{\NHI}{$N_\mathrm{H\kern 0.2em\textsc{i}}$}
\newcommand{\mNHI}{N_\mathrm{H\kern 0.2em\textsc{i}}}
\newcommand{\mNSiIII}{N_\mathrm{Si\kern 0.2em\textsc{iii}}}
\newcommand{\mNSiIV}{N_\mathrm{Si\kern 0.2em\textsc{iv}}}

\newcommand{\HI}   {{\rm H}{\sc \,i}}
\newcommand{\HeII} {{\rm He}{\sc \,ii}}
\newcommand{\FeIII} {{\rm Fe}{\sc \,iii}}
\newcommand{\CII}  {{\rm C}{\sc \,ii}}

\newcommand{\NV}  {{\rm N}{\sc \,v}}
\newcommand{\OI}   {{\rm O}{\sc\,i}}
\newcommand{\OVI}  {{\rm O}{\sc\,vi}}
\newcommand{\CIII} {{\rm C}{\sc \,iii}}
\newcommand{\CIV}  {{\rm C}{\sc \,iv}}
\newcommand{\AlII} {{\rm Al}{\sc \,ii}}
\newcommand{\AlIII}{{\rm Al}{\sc \,iii}}
\newcommand{\SiII} {{\rm Si}{\sc \,ii}}
\newcommand{\SiIII}{{\rm Si}{\sc \,iii}}
\newcommand{\SiIV} {{\rm Si}{\sc \,iv}}

\newcommand{\beginsupplement}{%
  \setcounter{table}{0}
  \renewcommand{\thetable}{S\arabic{table}}%
  \setcounter{figure}{0}
  \renewcommand{\thefigure}{S\arabic{figure}}%
}

\title[Possible Population III remnants at z=3.5]{Possible Population III Remnants at Redshift 3.5}

\author[N. H. M. Crighton et al.]{
Neil H. M. Crighton,$^1$\thanks{E-mail: neilcrighton@gmail.com}
John M. O'Meara$^2$
and Michael T. Murphy$^1$
\\
$^1$Centre for Astrophysics and Supercomputing, Swinburne University of Technology, Hawthorn, Victoria 3122, Australia\\
$^2$Department of Chemistry \& Physics, Saint Michael's  College, One Winooski Park, Colchester VT, 05439\\
}
\date{To appear in MNRAS letters}

\begin{document}
\label{firstpage}
\pagerange{\pageref{firstpage}--\pageref{lastpage}}
\maketitle

\begin{abstract}
The first stars, known as Population III (PopIII), produced the first
heavy elements, thereby enriching their surrounding pristine
gas. Previous detections of metals in intergalactic gas clouds,
however, find a heavy element enrichment larger than $1/1000$ times
that of the solar environment, higher than expected for PopIII
remnants. In this letter we report the discovery of a Lyman limit
system (LLS) at $z=3.53$ with the lowest metallicity seen in gas with
discernable metals, $10^{-3.41\pm0.26}$ times the solar value, at a
level expected for PopIII remnants. We make the first relative
abundance measurement in such low metallicity gas: the
carbon-to-silicon ratio is $10^{-0.26\pm0.17}$ times the solar
value. This is consistent with models of gas enrichment by a PopIII
star formation event early in the Universe, but also consistent with
later, Population II enrichment. The metals in all three components
comprising the LLS, which has a velocity width of 400~\kms, are offset
in velocity by $\sim+6\,$\kms\ from the bulk of the hydrogen,
suggesting the LLS was enriched by a single event.  Relative abundance
measurements in this near-pristine regime open a new avenue for
testing models of early gas enrichment and metal mixing.
\end{abstract}




\section{Introduction}

In the first three minutes after the Big Bang, nucleosynthesis
determined the hydrogen-to-helium ratio of pristine gas, which
contained no elements heavier than beryllium. The first heavier
elements (`metals') were manufactured by PopIII stars, and their
explosions polluted this pristine gas with metals
\citep{Yoshida04,Ciardi05}. The amount of pristine gas polluted and
the relative abundances of different heavy elements encodes
information about the mass distribution, nucleosynthetic processes and
other characteristics of the first stars \citep{Heger10,Bromm13}. The
recent discovery of gas without any observable metals, two billion
years after the big bang \citep{Fumagalli11_sci}, has opened the
possibility of finding new clouds enriched to the levels expected for
PopIII remnants. However, despite targeted searches
\citep[e.g.][]{Cooke11} for metal-poor systems, all gas clouds where
metals have been detected are found to be enriched to at least 1/1000
the solar value, higher than expected for gas enriched by the first
stars.

Here we report the discovery of a gas cloud with metallicity
$\sim1/2500$ solar, found by searching through archived observations
of quasars made using the UVES echelle spectrograph on the ESO Very
Large Telescope.

\vspace{-0.3cm}
\section{Data}

The cloud is part of an absorption system towards the background
quasar SDSS J124957.23$-$015928.8 ($z_\mathrm{em}=3.634$,
\citealt{Schneider03}). We identified the system by its optically
thick \HI\ absorption at the Lyman limit
($\lambda_\mathrm{rest}=912\,$\AA), which attenuates the quasar
continuum at observed wavelengths shorter than
$4140$\AA\ (Fig.~\ref{f:LL}). This Lyman limit system (LLS),
henceforth LLS1249, has a redshift $z=3.53$, corresponding to 1.8
billion years after the big bang (assuming a cosmology found by the
Planck Collaboration, \citeyear{Planck15_XIII}).

The archived UVES spectra of SDSS J124957.23 $-$015928.8 were taken
for program 075.A$-$0464 (P.I. Kim), and cover a wavelength range
$3750$--$6800\,$\AA\ at a resolution of $6\,$\kms\ full width at half
maximum (FWHM). Our analysis also uses an archived HIRES spectrum
($4110-8675\,$\AA, FWHM $6\,$\kms) of the quasar taken at the Keck
telescope during program U157Hb (P.I. Prochaska), and a Sloan Digital
Sky Survey (SDSS) spectrum with FWHM $\sim150\,$\kms.

The UVES spectra were reduced using the UVES pipeline, and combined
using the \textsc{uves\_popler} code.  A detailed explanation of the
reduction procedure is provided by, for example,
\citet{Bagdonaite12}. The HIRES spectrum was reduced using the
\textsc{HIRESredux} code as described by \citet{OMeara15}. For all
spectra we estimated the continuum over the \Lya\ forest using a
smoothly varying spline function, fitted by eye with interactive
tools.

\vspace{-0.3cm}
\section{Analysis}

\begin{figure*}
\includegraphics[width=0.75\textwidth,valign=t]{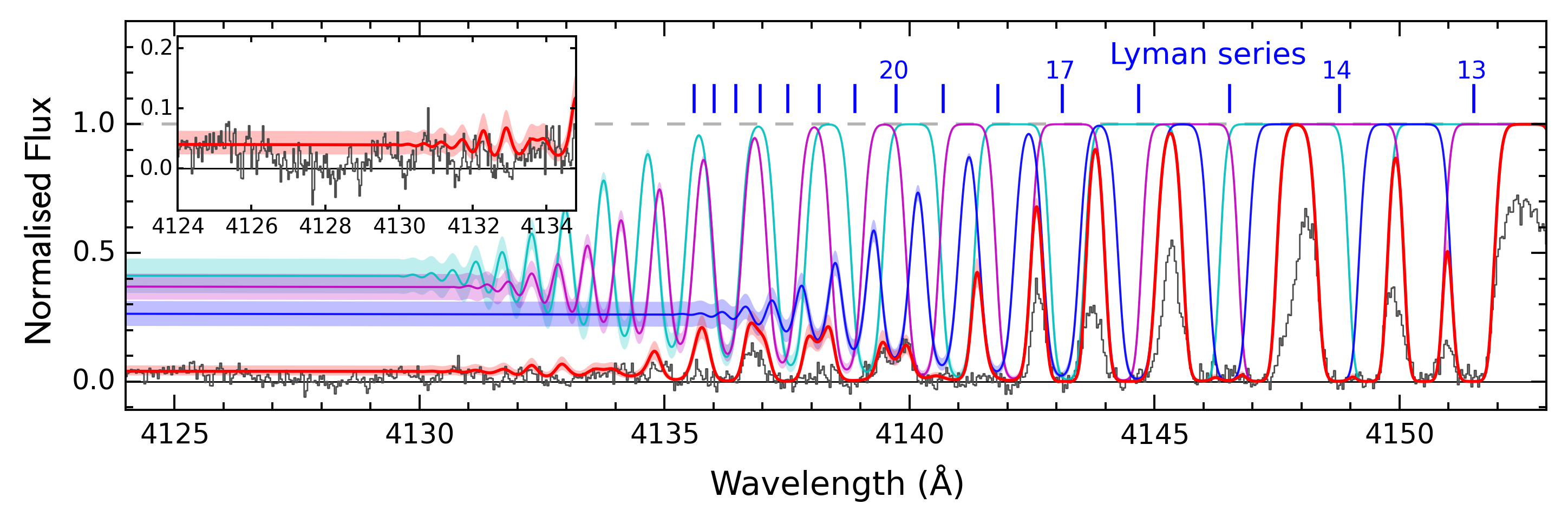}
\includegraphics[width=0.24\textwidth,valign=t]{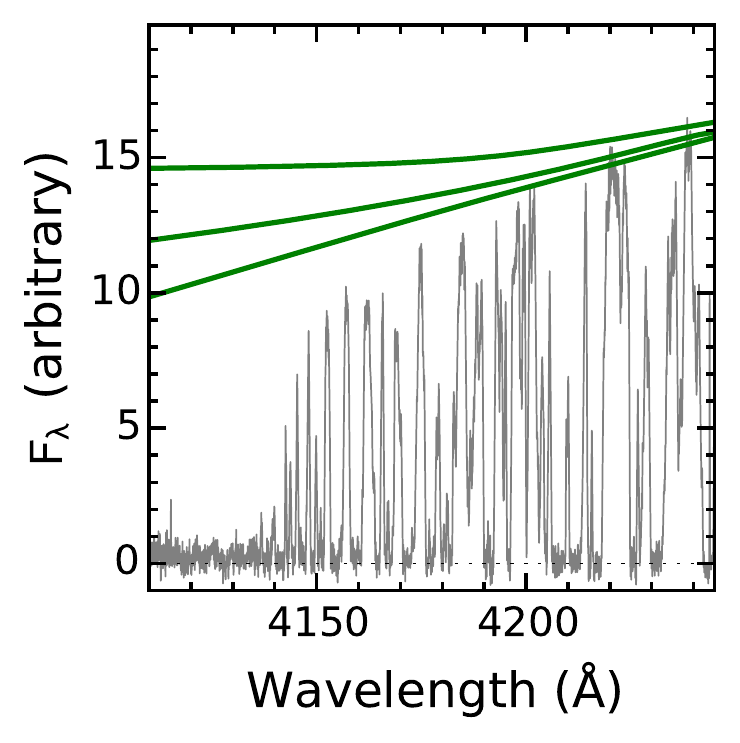}
\vspace{-0.3cm}
\caption{\textit{Left:} Lyman limit for LLS1249 in the spectrum of
  SDSS J124957.23$-$015928.8. The quasar spectrum divided by the
  continuum level is shown in black. Cyan, magenta and blue lines show
  our model for the three absorption components comprising LLS1249,
  and red shows the combined absorption model. These components
  reproduce both the higher order Lyman series absorption at
  4134$-$4152~\AA, marked by ticks for component 3 (blue), and the
  drop in flux at the Lyman limit below 4131~\AA. Shading shows
  2$\sigma$ errors on \NHI. \textit{Right:} The un-normalised UVES
  spectrum and the best (middle), highest (top) and lowest (bottom)
  continuum levels we adopt over the Lyman limit. We estimate the
  systematic error in \NHI\ associated with continuum estimation as
  half the difference between \NHI\ values found using the highest and
  lowest continua, and add this in quadrature to the statistical
  errors from our Voigt profile fits. The continua over the entire
  spectrum and for metal transitions of interest are shown in the
  online supplemental material in Fig.~S1--S7.
\label{f:LL}
}
\vspace{-0.4cm}
\end{figure*}

\begin{figure}
\begin{center}
\includegraphics[width=0.45\textwidth]{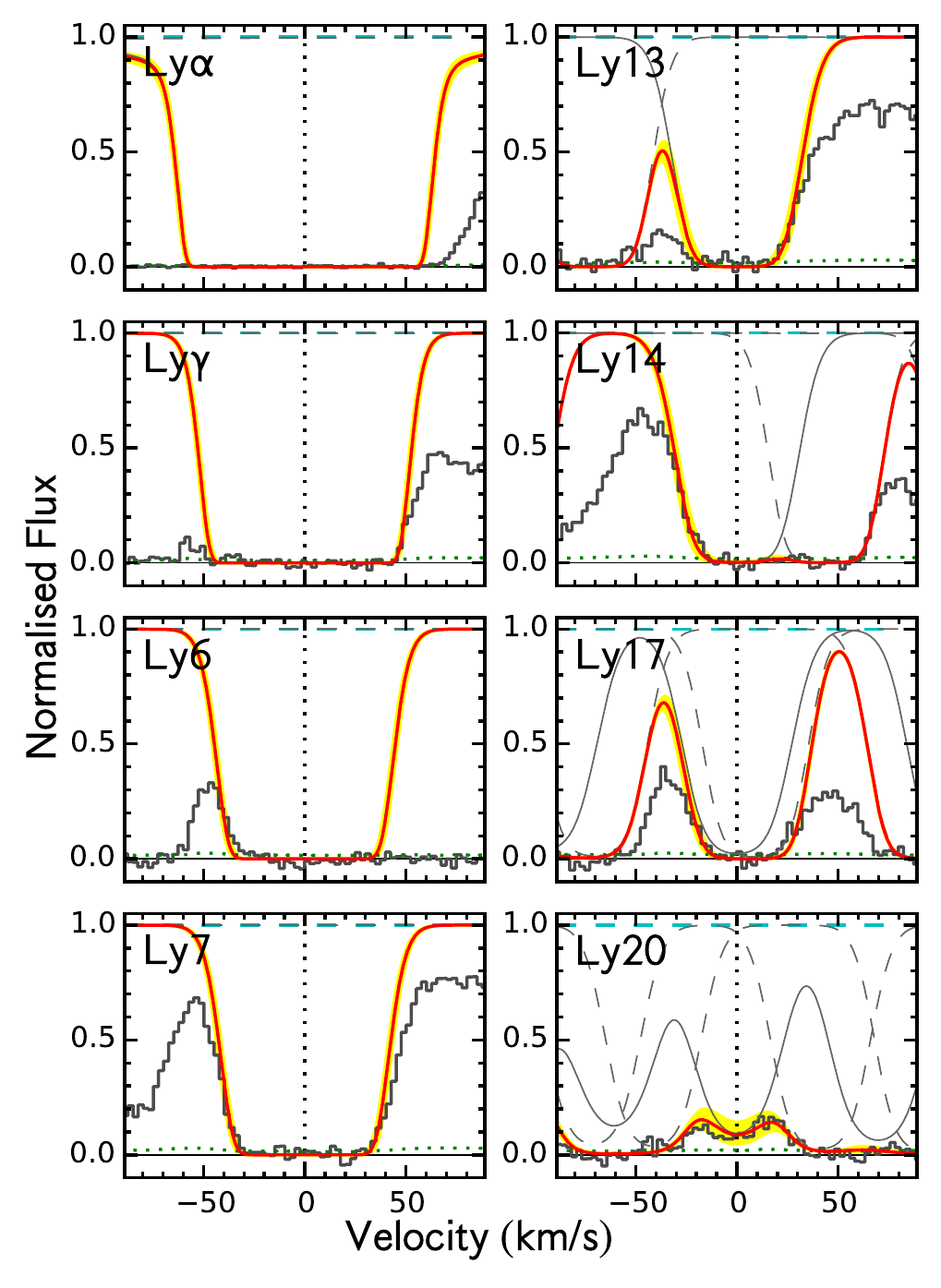}
\vspace{-0.3cm}
\caption{Lyman series for \HI\ component 3. The black histogram shows
  the quasar spectrum divided by the continuum, and yellow shows the
  $3\sigma$ errors.  Absorption from component 3 is shown as a thin
  solid line, and dashed lines are absorption from the two other
  components. The thicker red curve shows the combined absorption for
  all three components, and the dotted line near zero flux is the
  $1\sigma$ uncertainty in the flux. Zero velocity is at the redshift
  of component 3, $z=3.530073$. The column density, redshift and $b$
  parameter for this component are well determined by multiple Lyman
  series transitions.
}
\label{f:HI}
\vspace{-0.5cm}
\end{center}
\end{figure}

The Lyman series in the UVES combined spectrum reveals three
\HI\ absorption components spanning 400~\kms\ (Fig.~\ref{f:LL}). The
ratio between the residual flux and the continuum at the Lyman limit
sets the total \HI\ column density, \NHI, for these components, and
their Lyman series determine how \NHI\ is shared among them.  We fit
Voigt profiles to the three components to find their linewidths,
redshifts and \NHI. Table~\ref{t:NHI} lists the best-fitting
\HI\ parameters, and Fig.~\ref{f:HI} shows the Lyman series
transitions which best constrain \NHI\ for the highest redshift
component, component 3. We checked our \HI\ model by comparing it to
the SDSS spectrum, and find that it provides a good match to the
partial Lyman limit and the Lyman series in the lower resolution
spectrum.

\begin{table}
\addtolength{\tabcolsep}{-3pt}
\begin{center}
\caption{\label{t:NHI} The redshifts, column densities, and $b$ values
  for the three \HI\ components. The \NHI\ errors include a
  contribution from uncertainties in placing the continuum
  (see the right panel of Fig.~\ref{f:LL}).}
\begin{tabular}{ccccc}
\hline Comp. & $z$ & $\delta v$ (\kms)&
$\log(\mNHI/\mathrm{cm^{-2}})$ & $b$ (\kms) \\ \hline 1 &
$3.524038\,(04)$ & $-400$ & $17.15 \pm 0.04$ & $18.9 \pm 0.3$ \\ 2 &
$3.525285\,(12)$ & $-317$ & $17.20 \pm 0.03$ & $19.2 \pm 0.9$ \\ 3 &
$3.530073\,(04)$ & 0      & $17.33 \pm 0.03$ & $20.9 \pm 0.3$ \\ \hline
\end{tabular}
\end{center}
\vspace{-0.5cm}
\end{table}

The strongest metal transitions detectable for LLSs, which are
invariably highly ionized, are those of single, double and
triple-ionized carbon and silicon (\CII\ 1334, \CIII\ 977,
\CIV\ 1548/1550 and \SiII\ 1260, \SiIII\ 1206, \SiIV\ 1393/1402).
\CIII\ is detected in all three components, but only component 3 has
multiple metal transitions detected, so it affords the tightest
constraint on the ionization state and thus the gas metallicity.

\begin{figure}
\begin{center}
\includegraphics[width=0.42\textwidth]{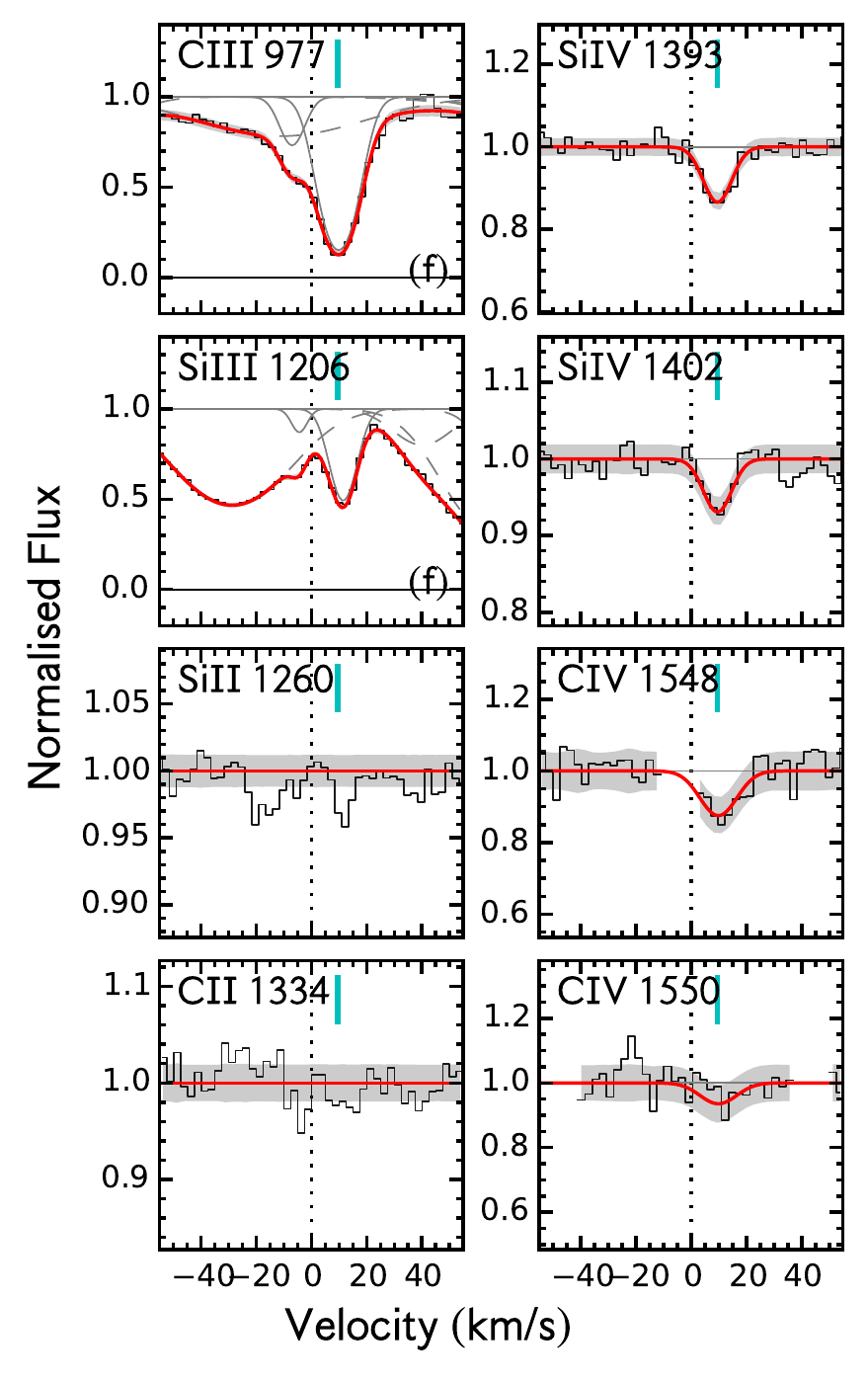}
\vspace{-0.5cm}
\caption{Metal transitions for component 3, where we measure the
  metallicity and carbon-to-silicon ratio. The cyan vertical ticks
  show metals from component 3. Grey shaded regions show the $1
  \sigma$ errors in the flux and the red line shows the combined Voigt
  profile fits. `(f)' means the transition is inside the \Lya\ forest,
  and is blended with unrelated \HI\ lines which are shown by dashed
  profiles. The zero velocity is the \HI\ component redshift.  Note
  the y limits change across different subpanels.}
\label{f:metals}
\vspace{-0.5cm}
\end{center}
\end{figure}

Fig.~\ref{f:metals} shows the metal transitions in component 3, and
Table~\ref{t:Nmet} lists the inferred column densities and $b$
parameters, for line detections. The line widths of the silicon and
carbon transitions are each consistent with a single value, suggesting
that both the low (\SiIII, \CIII) and high ionization (\SiIV, \CIV)
transitions are produced by the same gas phase.  Previous work
\citep{Prochaska99_Z_LLS,DOdorico01_LLS,Lehner14_OVI} has shown that
LLSs can be successfully modelled as gas clouds in photoionization
equilibrium with an ambient UV radiation field produced by the
integrated emission from galaxies and quasars
\citep[e.g.][HM12]{Haardt12}. In this context, the ionization state of
a cloud is determined by the ionization parameter $U$, the ratio of
the ionizing photon density to the gas density. Assuming the gas is in
photoionization equilibrium, the ratio $\mNSiIII/\mNSiIV$ sets a lower
limit $U>10^{-3}$.\footnote{We note that this validates Fumagalli et
  al.'s (\citeyear{Fumagalli11_sci}) assumption that $U>10^{-3}$ in
  very low metallicity LLS.} This corresponds to a gas volume density
$n_\mathrm{H} < 10^{-2}$ and H neutral fraction $x_\mathrm{H}<0.04$. The
thickness of the cloud is given by $\mNHI/(x_\mathrm{H} n_\mathrm{H})$. We
can estimate an upper limit to the cloud size from the observed width
of the \HI\ line for component 3, which has
$b=20.9\pm0.3\,$\kms\ (corresponding to a full width at half maximum,
FWHM $\sim35\,$\kms). If the cloud is large enough it will expand with
the cosmological Hubble flow, producing a velocity gradient which will
be detectable in the absorption profile
\citep{Simcoe02,Levshakov03_lowZ}. The resulting line broadening is
given by the Hubble parameter $H(z=3.53)=375\,$\kms$\,$Mpc$^{-1}$, and
thus the largest size allowed by the broadening observed is
approximately FWHM$/(375\,$\kms$\,$Mpc$^{-1})=90$~physical kpc. This
translates to an upper limit $U<10^{-2}$, where we have included a
factor of 2.2 uncertainty in the normalization of the incident
radiation field in converting between the cloud volume density and the
$U$ value \citep[e.g.][]{Becker13_gamma}. The metallicity range
corresponding to these $U$ limits is $10^{-4}<Z/Z_\odot<10^{-2.6}$,
where we assume $\log_{10}(Z/Z_\odot)=~$[Si/H]. If we assume a larger uncertainty
in the radiation field normalization, this translates to a lower
limiting metallicity.

\begin{table}
\renewcommand{\arraystretch}{1.15}
\addtolength{\tabcolsep}{-3.6pt}
\footnotesize
\begin{center}
\caption{\label{t:Nmet} Parameters inferred for metals in the three
  components. Column densities are measured using the apparent optical
  depth method; Voigt profile fitting gives a consistent result. Upper
  limits are $3\sigma$ and assume an optically thin
  transition. Uncertainties include a 5\% systematic uncertainty in
  the continuum level (10\% inside the \Lya\ forest), and an
  uncertainty in the zero level of 0.5$\sigma_\mathrm{flux}$.}
\begin{tabular}{rcrcrc}
\multicolumn{6}{c}{Upper limits}\\
\multicolumn{2}{c}{Comp. 1} & \multicolumn{2}{c}{Comp. 2} & \multicolumn{2}{c}{Comp. 3} \\ 
Ion & $\log (N/\mathrm{cm^{-2}})$ & Ion & $\log (N/\mathrm{cm^{-2}})$ & Ion & $\log (N/\mathrm{cm^{-2}})$ \\
\hline
 \AlII  & $<10.94   $    & \AlII & $<11.31   $  & \AlII &   $<11.31   $ \\
 \AlIII & $<11.81   $    & \AlIII& $<12.10   $  & \AlIII&   $<12.10   $ \\
 \CII   & $<11.65   $    & \CII  & $<12.42   $  & \CII  &   $<12.42   $ \\
 \NV    & $<12.47    $   & \CIV  &  $<12.64   $ & \NV   &  $<12.89    $ \\
 \OI    & $<12.08    $   & \NV   &  $<12.89    $& \OI   &  $<12.44    $ \\
 \OVI   & $<12.84    $   & \OI   &  $<12.44    $& \OVI  &  $<13.17    $ \\
 \FeIII & $<12.07    $   & \OVI  &  $<13.17    $& \FeIII&  $<12.74    $ \\
 \SiII  & $<10.50    $   & \FeIII&  $<12.74    $&       &  \\
 \SiIV  & $<11.05$       & \SiII &  $<11.25    $&       &  \\
        &                & \SiIII&  $<12.26    $&       &  \\
        &                & \SiIV &  $<12.30    $&       &  \\
\hline
\end{tabular}
\begin{tabular}{crccc}
\multicolumn{5}{c}{Detections}\\
Comp. & Ion & $\log (N/\mathrm{cm^{-2}})$ & $b$ (\kms) & $z$\\
\hline
1 & \CIII        &$12.52_{-0.13}^{+0.13}$& $10.0\pm0.5$   & $3.524168(06)$ \\
1 & \CIV         &$12.44_{-0.13}^{+0.13}$&$12.0\pm4.7$    & $3.524168^\mathrm{b}$ \\
2 & \CIII        &$12.14_{-0.18}^{+0.25}$& $9.1\pm1.1$    & $3.525391(10)$ \\
3 & \SiII        &$10.96_{-0.33}^{+0.28}$& $5.4^\mathrm{c}$  & $3.530247^\mathrm{c}$ \\
3 & \SiIII       &$12.20_{-0.15}^{+0.15}$& $5.4\pm0.3$    & $3.530247(02)$\\
3 & \SiIV        &$12.04_{-0.15}^{+0.13}$& $5.6\pm0.6$	& $3.530217(05)$\\
3 & \CIII        &$13.17_{-0.04}^{+0.05}$& $7.7\pm0.4$    & $3.530222(04)$\\
3 & \CIV$^\mathrm{a}$ &$12.45_{-0.08}^{+0.08}$&$9.0\pm2.2$    & $3.530222^\mathrm{b}$\\
\hline
\end{tabular}
\end{center}
$^\mathrm{a}$ \CIV\ 1548 is partly blended with sky emission.\\
$^\mathrm{b}$ Fixed at the \CIII\ redshift.\\
$^\mathrm{c}$ Fixed at at \SiIII\ redshift and $b$.\\
\vspace{-0.5cm}
\end{table}

To refine this metallicity measurement and determine other physical
properties of the cloud, we use Markov Chain Monte Carlo sampling to
compare the observed column densities to a grid of photoionization
models, including a variable slope for the incident radiation field,
$\alpha_\mathrm{UV}$ \citep{Crighton15_LAE}. In brief, a single phase,
constant density cloud is assumed, illuminated by the UV background
radiation from HM12. The \textsc{Cloudy} code
\citep[][v13.03]{Ferland13} is then used to create a grid of
photoionization models with predicted column densities to compare to
the observed values listed in Table \ref{t:Nmet}. Our grids cover a
range $-4.6 < \log n_\mathrm{\HI}/$\cmmm$ < -1.0$ (corresponding
to $-4.2 <\log U < -0.6$), $-2.5<\alpha_{UV}<1.5$ and $15.5<\log
\mNHI/$\cmm $<18.5$. Marginalizing over uncertainties in \NHI,
the radiation slope, and $U$ we find a 68\% range
$Z/Z_\odot=10^{-3.41\pm0.26}$, and a 95\% range
$10^{-3.82}<Z/Z_\odot<10^{-2.84}$.

\begin{figure}
\begin{center}
\includegraphics[width=0.5\textwidth]{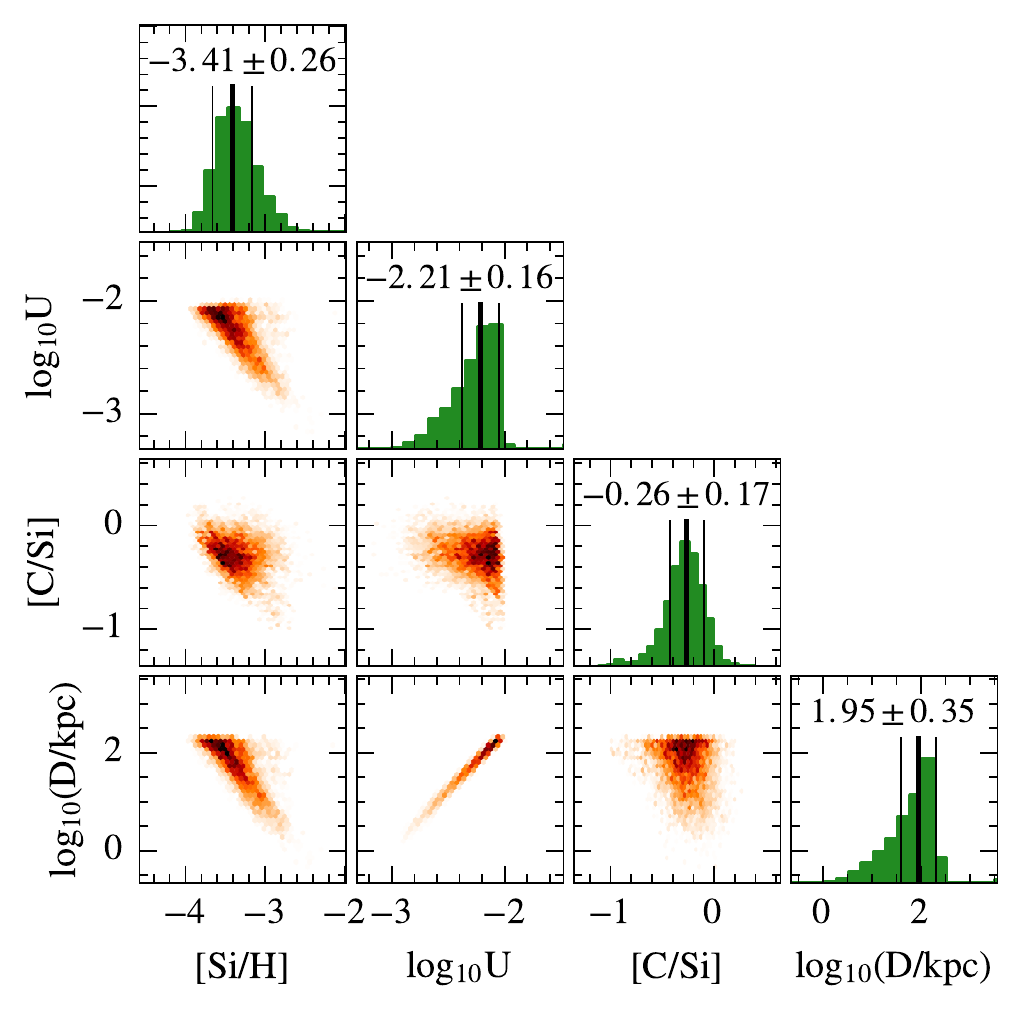}
\vspace{-0.4cm}
\caption{\label{f:par} Photoionization model parameters for component
  3, where we measure the metallicity and carbon-to-silicon
  ratio. Panels show the MCMC samples plotted as a function of
  parameter pairs taken from $U$, $\log_{10}$ of the metallicity
  relative to solar [Si/H], $\log_{10}$ of the carbon-to-silicon ratio
  relative to solar [C/Si], and the cloud size D. All parameters have
  flat priors apart from the cloud size, which was required to be
  $<200\,$kpc, and the radiation slope $\alpha_\mathrm{UV}$, as
  described in the text.}
\end{center}
\vspace{-0.5cm}
\end{figure}

Fig.~\ref{f:par} shows the parameters we derive for component 3. The
photoionization models imply a gas temperature $10^{4.35\pm0.05}\,$K,
consistent with the maximum temperature allowed by the \HI\ and metal
line widths, and $N_\mathrm{H}=10^{19.8}\mathrm{--}10^{20.8}\,$\cmm\ (95\%
confidence interval). The models favour an ionizing spectrum harder
than that in HM12. If no constraint is placed on $\alpha_\mathrm{UV}$,
the models prefer very hard spectra ($\alpha_\mathrm{UV}> 1$) which
are incompatible with the observed \HI\ and \HeII\ photoionization
rates \citep{McQuinn14}.  Crighton et al. (2015) showed this to be the
case at $z=2.5$, but it should also be true at the redshift of
LLS1249, $z=3.5$, particularly given the quasar space density drops by
a factor of 2$-$3 from $z=2.5$ to $3.5$, which should result in an
even softer spectrum at the higher redshift. Therefore we applied a
Gaussian prior on $\alpha_\mathrm{UV}$ with $\sigma=0.5$ centred on
0. This provides a satisfactory fit to the column densities, while
still matching the \HeII\ observational constraints. If we require the
slope to match the HM12 spectrum precisely
(i.e. $\alpha_\mathrm{UV}=0$) this results in an even lower
metallicity.

Motivated by the small velocity shifts between the \HI\ and metals,
described in the following section, we also explore a two component
photoionization model in which the metals are not associated with the
bulk of the \HI. To test this scenario, we ran further ionization
models, splitting component 3 into two phases which we allow to have
different $U$. We assume that all of the metal lines are produced in
one phase, and require that the amount of \HI\ in this phase is less
than the \NHI\ measured for the whole component. This leads to a
solution where the metals are produced by a sub-component with \NHI
$\,< 10^{16}\,$\cmm, $U \sim 10^{-2.3}$ and metallicity $\sim0.01$
solar. In this case most of the \HI\ would be associated with a second
phase that does not have detectable metals. Without any metals
detected we are unable to derive a $U$ value, but using the upper
limits on metals from the spectrum, and assuming $U>10^{-3}$
\citep{Fumagalli11_sci}, we find an upper limit on the metallicity of
$Z/Z_\odot<10^{-3.8}$. Therefore this two phase scenario sees a pocket
of enriched gas embedded in a surrounding pristine cloud. If this two
phase model is correct, our interpretation of the absorbing cloud is
unchanged. Indeed, this is entirely consistent with the two scenarios
we consider in the next section, which both involve a pristine gas
cloud polluted by smaller clumps of more metal-enriched gas.

Finally, we checked that the metallicity of LLS1249 does not not
depend strongly on the uncertainty in our measured \NHI. If we assume
an uncertainty in \NHI\ for component 3 of 0.2 dex, more than six
times larger than our estimated error in Table~\ref{t:NHI}, the
resulting metallicity is $10^{-3.38\pm0.26}$, close to our best estimate
of $10^{-3.41\pm0.26}$.

\section{Discussion}

\begin{figure}
\begin{center}
\includegraphics[width=0.45\textwidth]{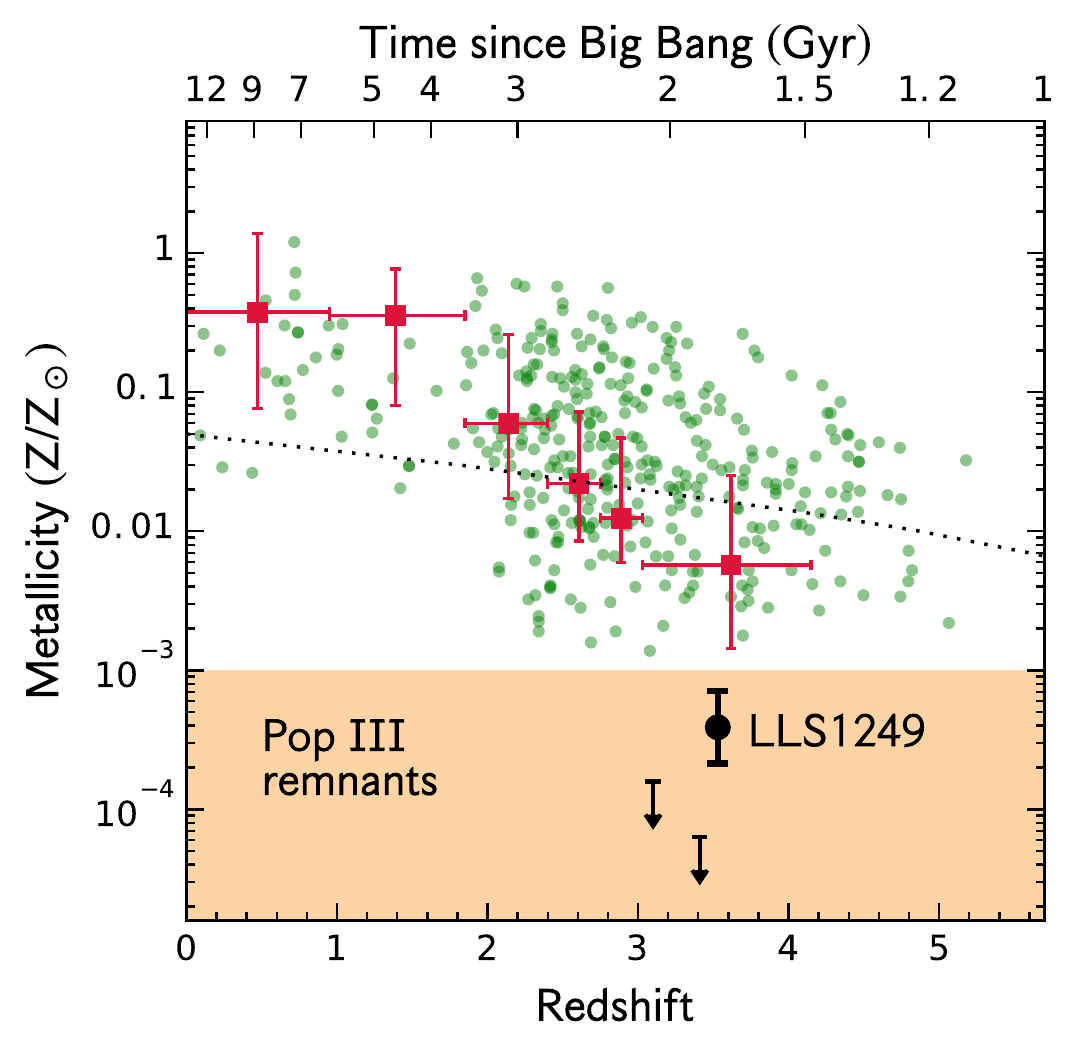}
\vspace{-0.4cm}
\end{center}
\caption{The metallicity of LLS1249 compared to other measurements of
  diffuse gas metallicities. Green dots show representative
  metallicities for the predominantly neutral damped Lyman-$\alpha$
  systems (see text); their typical uncertainties are 0.2 dex. Red
  squares show metallicities for a compilation of the more highly
  ionized Lyman limit systems \citep{Fumagalli15}. Vertical errors on
  the squares show the 25--75\% range in the sample metallicity
  probability distribution.  Upper limits from \citet{Fumagalli11_sci}
  are shown as arrows. The large dot with error bars shows our
  measurement in LLS1249 and its $1\sigma$ uncertainty. The dotted
  line shows an analytic model for the cosmological evolution of gas
  metallicity \citep{Hernquist03}. The shaded region shows predictions
  for gas metallicities resulting from PopIII star formation events
  \citep{Wise08,Wise12}. \vspace{-0.4cm}}
\label{f:cf_Z}
\vspace{-0.1cm}
\end{figure}

Fig.~\ref{f:cf_Z} shows our result in comparison to previous
metallicity measurements in gas clouds. The metallicity of the diffuse
gas ($n_\mathrm{H}\sim10^{-4}\,$\cmmm, close to the cosmic mean
density) in the intergalactic medium is estimated at $\sim10^{-3.5}$
solar by statistical analyses of many weak absorption lines
\mbox{\citep{Schaye03,Simcoe11_Z}}.  Two Lyman limit systems have been
discovered without any detectable metals, giving upper limits of
$Z/Z_\odot<10^{-4.2}$ and $Z/Z_\odot<10^{-3.8}$
\citep{Fumagalli11_sci}. When metals have been detected in individual
LLS, which we loosely define here as having
$10^{17}\,$\cmm$<\mNHI<10^{20.3}\,$\cmm, they are found at a level
$Z/Z_\odot>10^{-3}$ (we show results from the recent compilation by
\citealt{Fumagalli15}, which includes most known low metallicity LLS,
see also \citealt{Cooper15}). Recent large samples of metallicity
measurements in damped \Lya\ systems (with $\mNHI>10^{20.3}\,$\cmm)
have also not detected metallicities lower than $10^{-3}\,$Z$_\odot$
\citep{Rafelski12,Jorgenson13}, even in targeted searches for
metal-poor systems \citep{Cooke11}. The metallicity of LLS1249 is thus
significantly below those measured in previous strong absorption
systems with detectable metals.

The photoionization models imply a carbon-to-silicon ratio
$10^{-0.26\pm0.17}$ solar for LLS1249. This ratio, together with the
overall very low metallicity, suggests two possible origins for the
absorbing cloud. The first is enrichment of a pristine cloud by the
earliest star formation events in the universe. In this scenario an
overdense, pristine region of the early universe collapses to produce
PopIII stars, and the death of these first stars curtails further
accretion by releasing energy larger than the binding energy of the
proto-galaxy. The pristine gas surrounding the proto-galaxy, now
polluted with PopIII remnants and unable to collapse further to form
new stars, expands with the Hubble flow, producing the absorption we
see. Supernovae from higher mass PopIII stars are able to eject their
metals to a large distance from the halo, consistent with this
scenario \citep{Cooke14_cemp}. The yields for these high-mass
supernovae predict a carbon-to-silicon ratio consistent with that
measured in LLS1249.

Components 1 and 2 in LLS1249 do not have sufficient metal transitions
to robustly determine $U$ values, but both do show
\CIII\ absorption. Assuming the $U$ for these components is similar to
component 3 implies they have a similarly low
metallicity. Interestingly, for all three components the metal
absorption is offset from the velocity of hydrogen by
$6$--$10\,$\kms\ (Fig.~\ref{f:metals} and online supplemental
Fig.~S8).  These shifts are consistent across the UVES and HIRES
spectra, taken on different telescopes and with wavelength
calibrations calculated by different software. Therefore they are
unlikely to be caused by wavelength calibration uncertainties. The
bulk of the metals thus have a slightly different velocity structure
to the \HI, implying they are not well mixed throughout the
cloud. Mixing timescales from diffusion and turbulence can be several
billions of years (Crighton et al., in preparation), so an
inhomogeneous metal distribution is consistent with a PopIII
enrichment scenario.  The offsets in every component have a similar
magnitude and direction, suggesting that a single enrichment event may
have been responsible for polluting the \HI\ gas. Assuming a spherical
cloud and the densities inferred from photoionization modelling, the
minimum mass for the gas producing component 3 is $10^8\,M_\odot$,
which implies a metal mass greater than
$10^{-3.4}\times0.014\times10^8\approx500\,M_\odot$. This is
significantly more than the metal yield of a single pair-instability
supernova, $\lesssim100\,M_\odot$\citep{Heger02}. Therefore if LLS1249 was
enriched by a single event it was likely a starburst consisting of at
least several supernovae, and not by an individual star. The recent
reported observation of a PopIII starburst at $z=6.6$ may represent
such an event \citep{Sobral15}.

Alternatively, the gas may have been enriched by recent star formation
at $z\lesssim4$. In this scenario pristine gas in either a cold
accretion stream \citep{Dekel09} or low-mass halo \citep{Cen08} is
polluted by Population II star formation from a nearby galaxy. We
consider a cold stream to be unlikely, however, because they are
expected to have a characteristic size of a few kpc and our
photoionization models favour much larger cloud sizes ($>6\,$kpc at
$95\%$ confidence). The cloud is also unlikely to be in a low-mass
halo, because the velocity width of LLS1249 is $>400\,$\kms, more than
seven times larger than the virial velocity of a halo with mass
$10^{10}\,$M$_\odot$ at $z=3.5$. The cloud may be illuminated by
ionizing radiation from a quasar, which would be consistent with the
hard ionizing spectrum the photoionization models prefer. If this
ionizing radiation is much stronger than the integrated UV background
we assume, this would also imply higher cloud densities and thus
smaller cloud sizes, possibly even consistent with a cold-accretion
stream. It would be surprising, however, to find such metal-poor gas
in the vicinity of a quasar, which are usually in dense environments
\citep{Shen07}.

New observations of extremely metal-poor absorption systems
\citep[e.g.][]{Cooper15} will help to clarify the origin of metals in
these systems. LLS with higher column densities
($\mNHI=10^{18\mathrm{--}20}\,$\cmm) may be discovered which, even at
these low metallicities, show absorption from aluminium, oxygen and
nitrogen, in addition to carbon and silicon. The O/Si abundance ratio
in particular is sensitive to the assumed PopIII initial mass
function \citep{Kulkarni13}, and will enable a more detailed
comparison to be made with models of pristine gas enrichment.

\vspace{-0.5cm}
\section*{Acknowledgements}

\vspace{-0.2cm}
 NHMC and MTM thank the Australian Research Council for
 \textsl{Discovery Project} grant DP130100568 which supported this
 work. Our analysis made use of \textsc{astropy} \citep{Astropy13},
 \textsc{uves\_popler}
 (\url{http://astronomy.swin.edu.au/~mmurphy/UVES_popler}),
 \textsc{HIRESredux} (\url{http://www.ucolick.org/~xavier/HIRedux}),
 \textsc{vpfit} (\url{http://www.ast.cam.ac.uk/~rfc/vpfit.html}), and
 \textsc{matplotlib} \citep{Hunter07}. This work used archived data
 taken at the ESO La Silla Paranal Observatory for program
 075.A$-$0464 (P.I. Kim) and at the W. M. Keck Observatory for program
 U157hb (P.I. Prochaska). The authors wish to acknowledge the
 significant cultural role that the summit of Maunakea has always had
 within the indigenous Hawaiian community.  We are most fortunate to
 have the opportunity to conduct observations from this mountain.





\beginsupplement

\vspace{1cm}

\bf{\Large Supplementary online material}


\begin{figure}
\begin{center}
\includegraphics[width=0.5\textwidth]{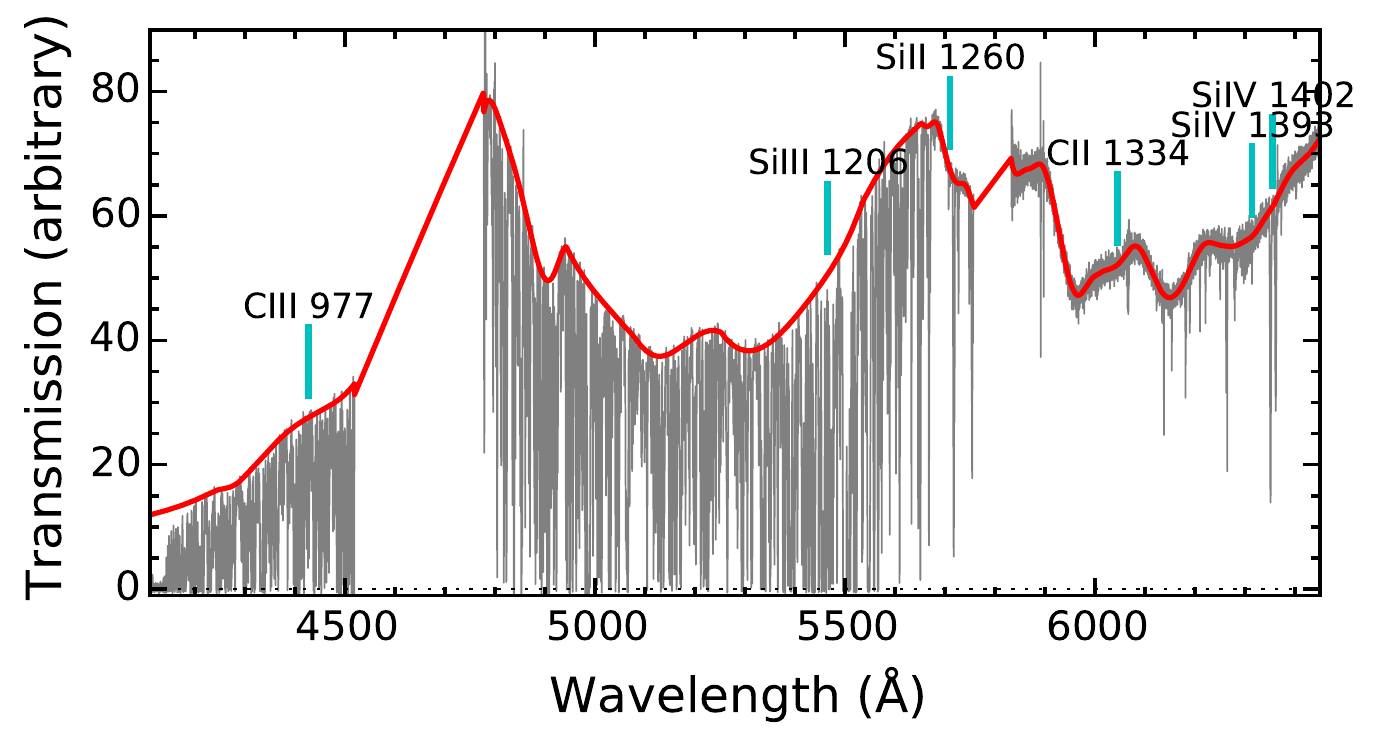}
\vspace{-0.3cm}
\caption{\label{f:co_all} The UVES spectrum used to measure
  \HI\ and metal line parameters. The adopted continuum is shown in
  red, and ticks show the positions of important metal transitions for
  component 3 in LLS1249. The QSO \Lya\ emission line is near
  5650\AA. The spectrum is not flux-calibrated, and the scaling of the
  transmission is arbitrary. The large gaps at 4500--4800 \AA\ and
  5750--5850 \AA\ are regions of the spectrum not covered by the UVES
  observations.}
\end{center}
\vspace{-0.3cm}
\end{figure}

\begin{figure}
\begin{center}
\includegraphics[width=0.5\textwidth]{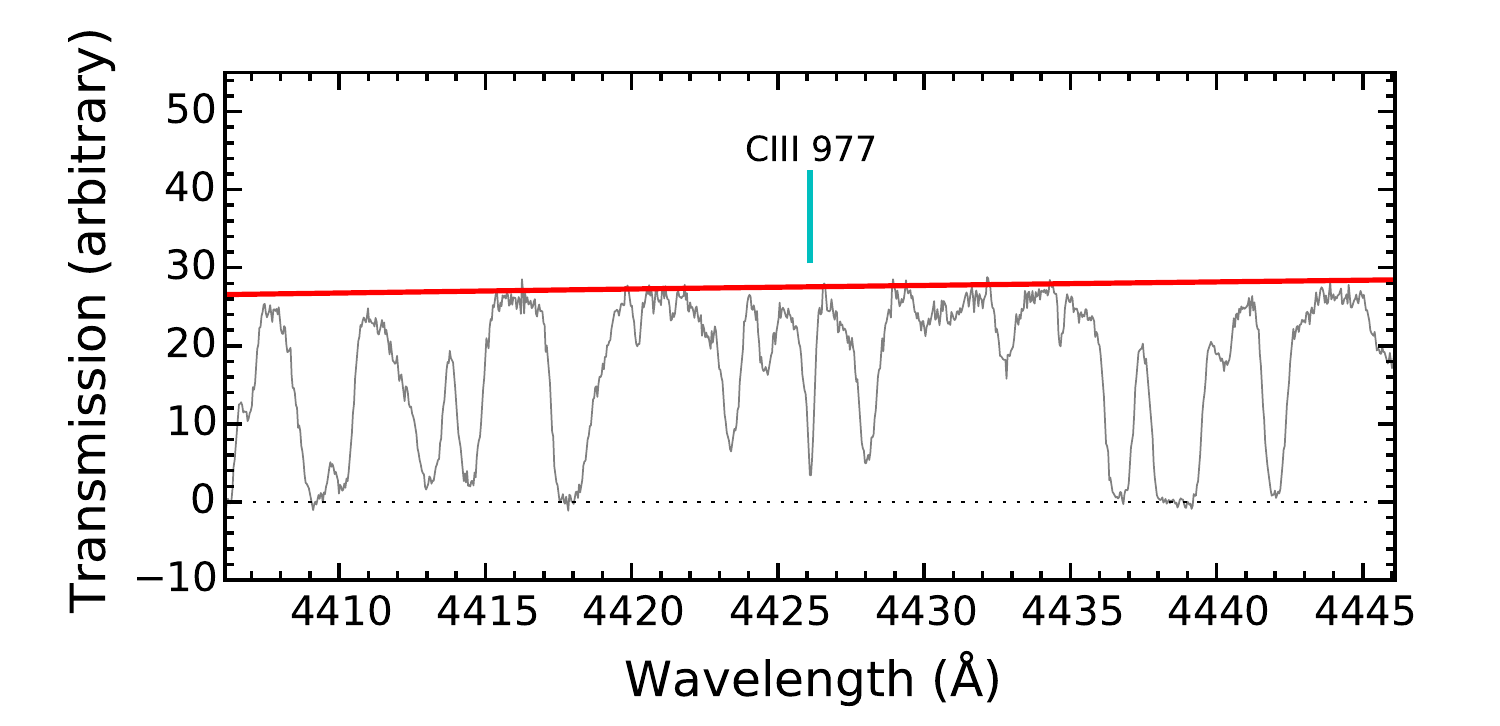}
\vspace{-0.3cm}
\caption{\label{f:CIII}  The continuum around \CIII\ 977 for component 3.}
\end{center}
\vspace{-0.3cm}
\end{figure}

\begin{figure}
\begin{center}
\includegraphics[width=0.5\textwidth]{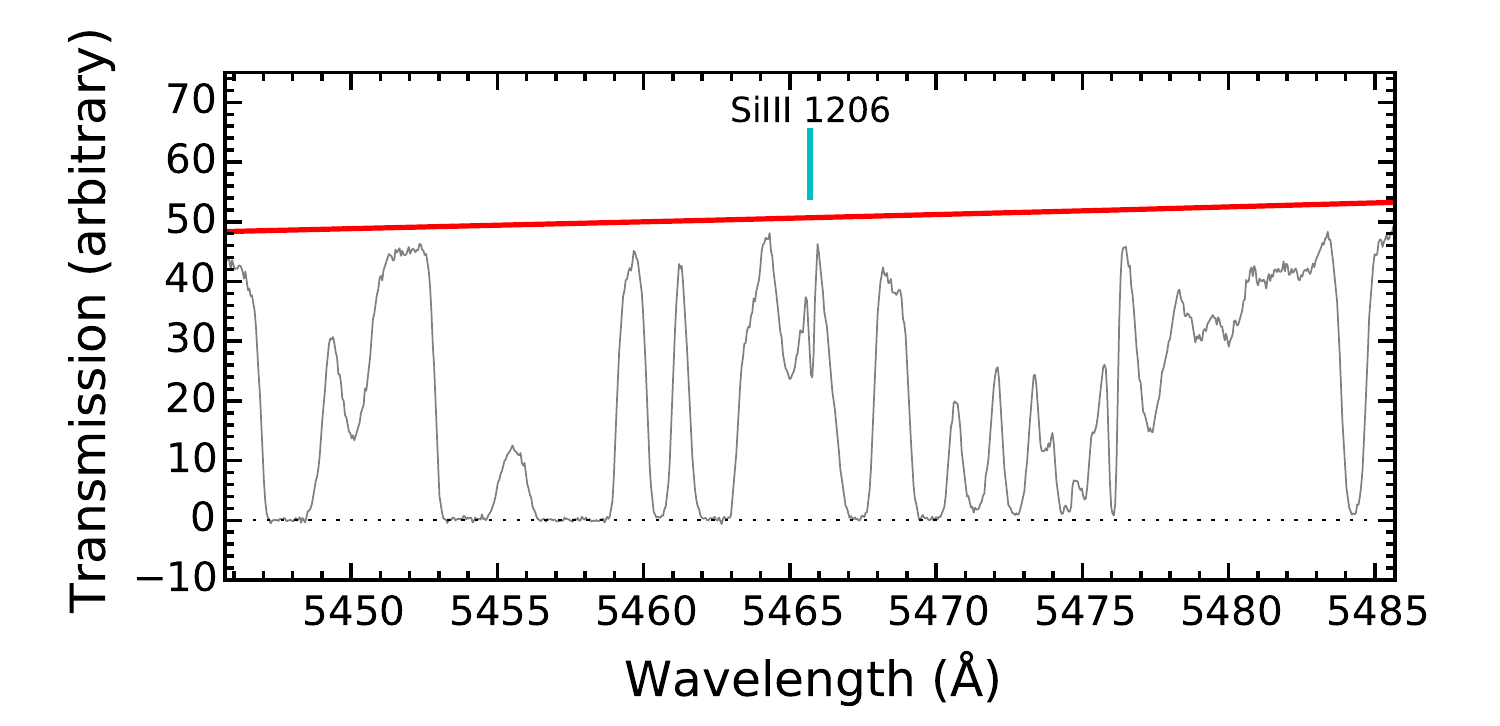}
\vspace{-0.3cm}
\caption{\label{f:SiIII} The continuum around \SiIII\ 1206 for
  component 3.}
\end{center}
\vspace{-0.3cm}
\end{figure}

\begin{figure}
\begin{center}
\includegraphics[width=0.5\textwidth]{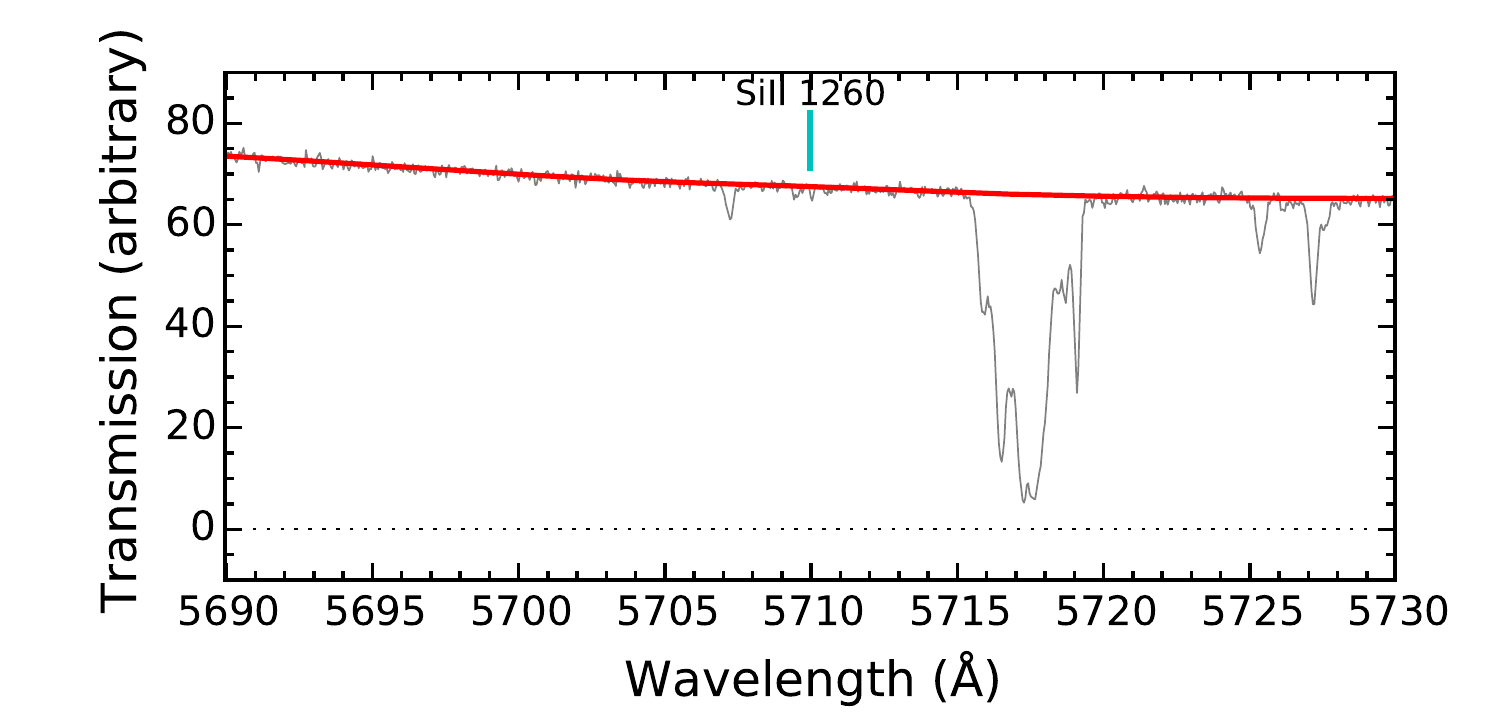}
\vspace{-0.3cm}
\caption{\label{f:SiII}  The continuum around \SiII\ 1260 for
  component 3.}
\end{center}
\vspace{-0.3cm}
\end{figure}

\begin{figure}
\begin{center}
\includegraphics[width=0.5\textwidth]{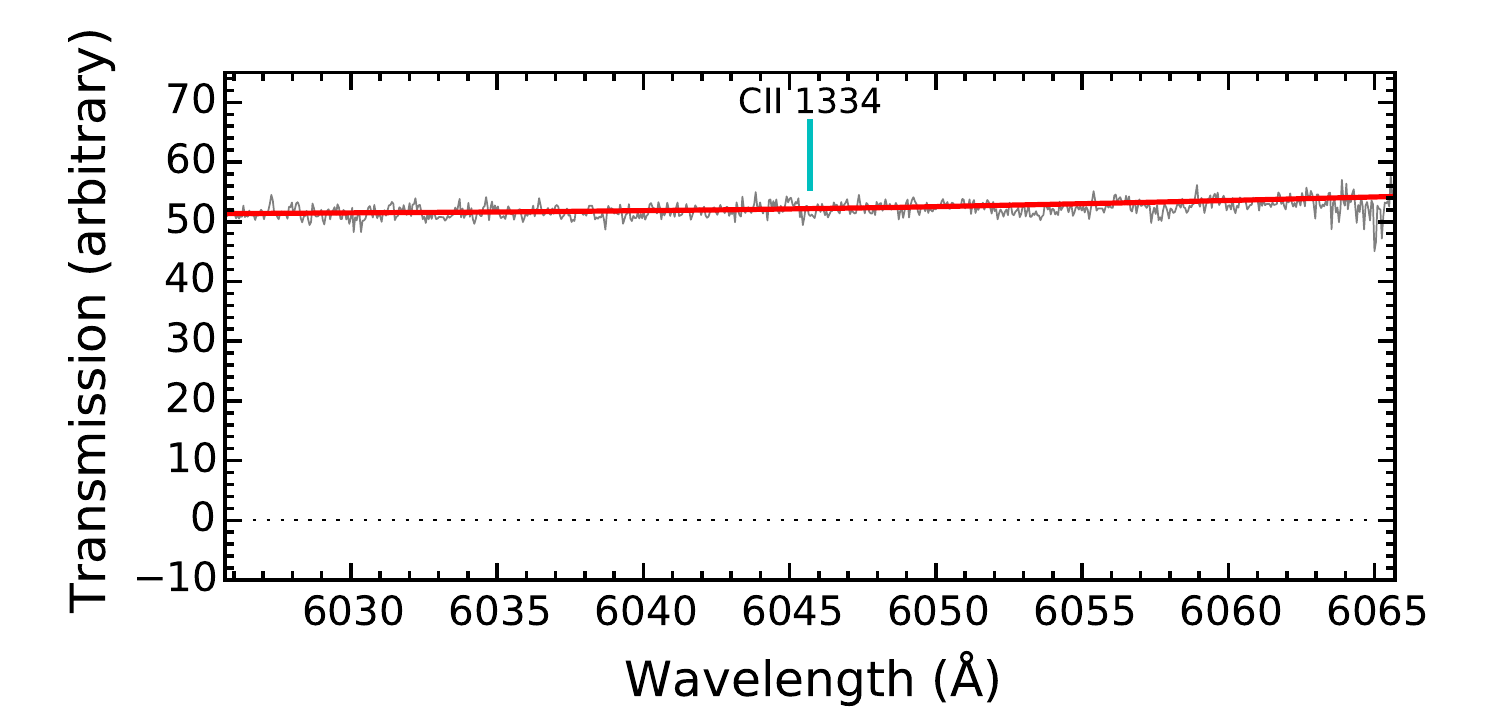}
\vspace{-0.3cm}
\caption{\label{f:CII}  The continuum around \CII\ 1334 for
  component 3.}
\end{center}
\vspace{-0.3cm}
\end{figure}

\begin{figure}
\begin{center}
\includegraphics[width=0.5\textwidth]{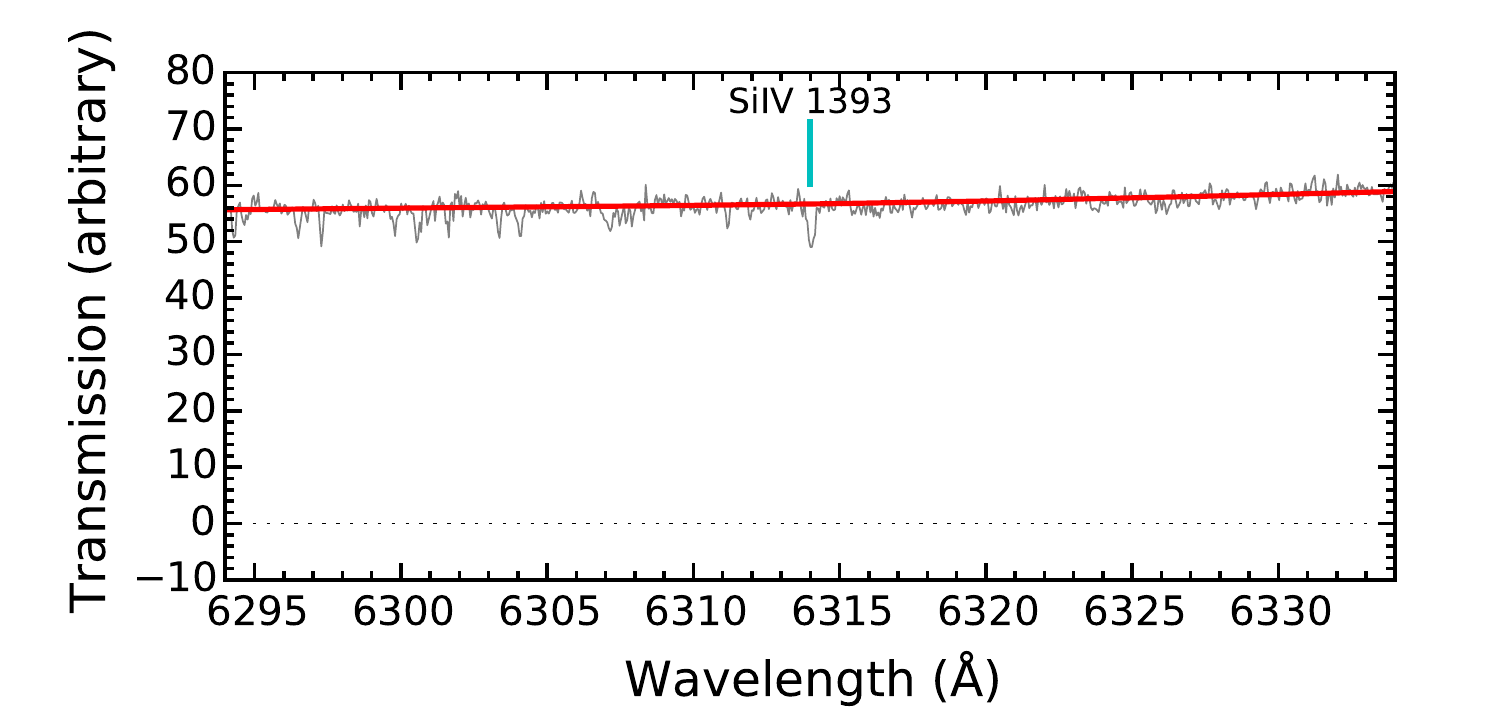}
\vspace{-0.3cm}
\caption{\label{f:SiIVa}  The continuum around \SiIV\ 1393 for
  component 3.}
\end{center}
\vspace{-0.3cm}
\end{figure}

\begin{figure}
\begin{center}
\includegraphics[width=0.5\textwidth]{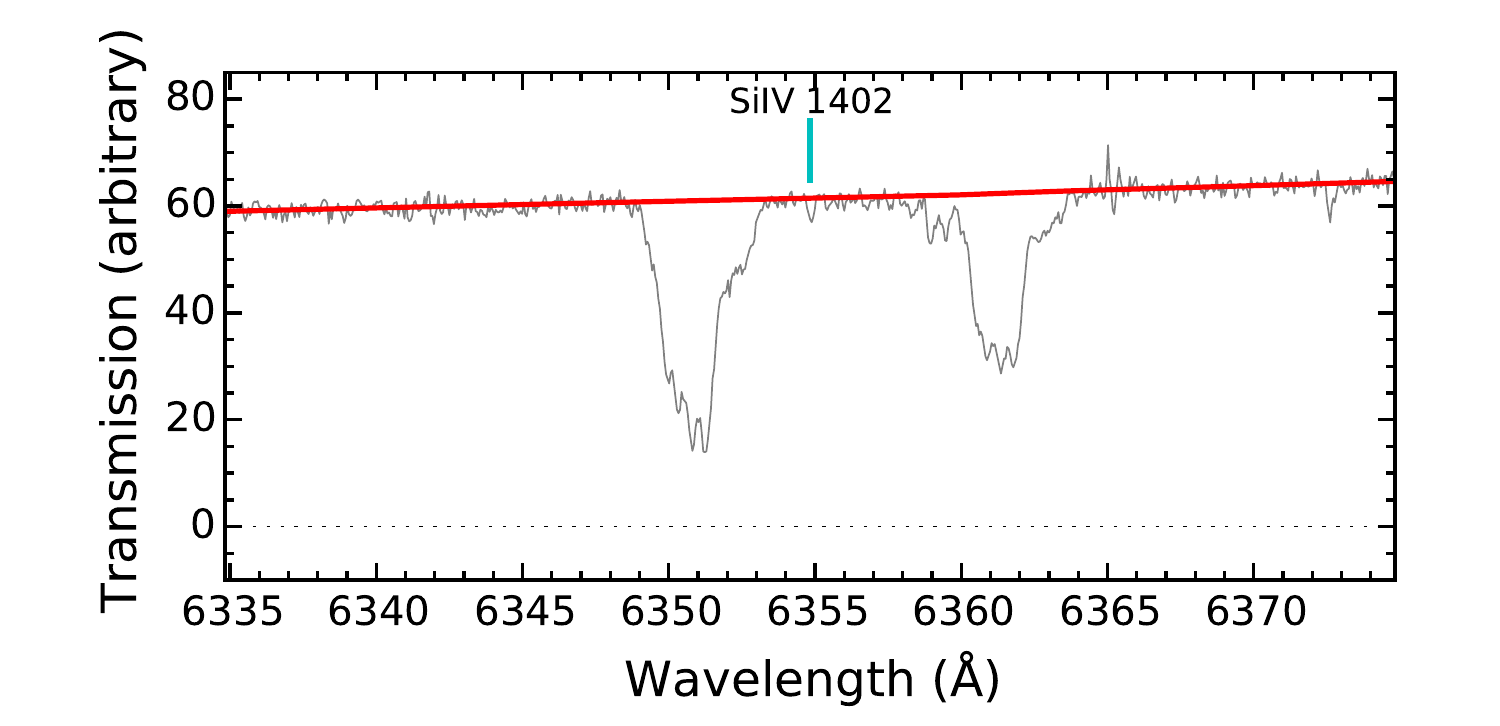}
\vspace{-0.3cm}
\caption{\label{f:SiIVb}  The continuum around \SiIV\ 1402 for
  component 3.}
\end{center}
\vspace{-0.3cm}
\end{figure}

\begin{figure}
\begin{center}
\includegraphics[width=0.5\textwidth]{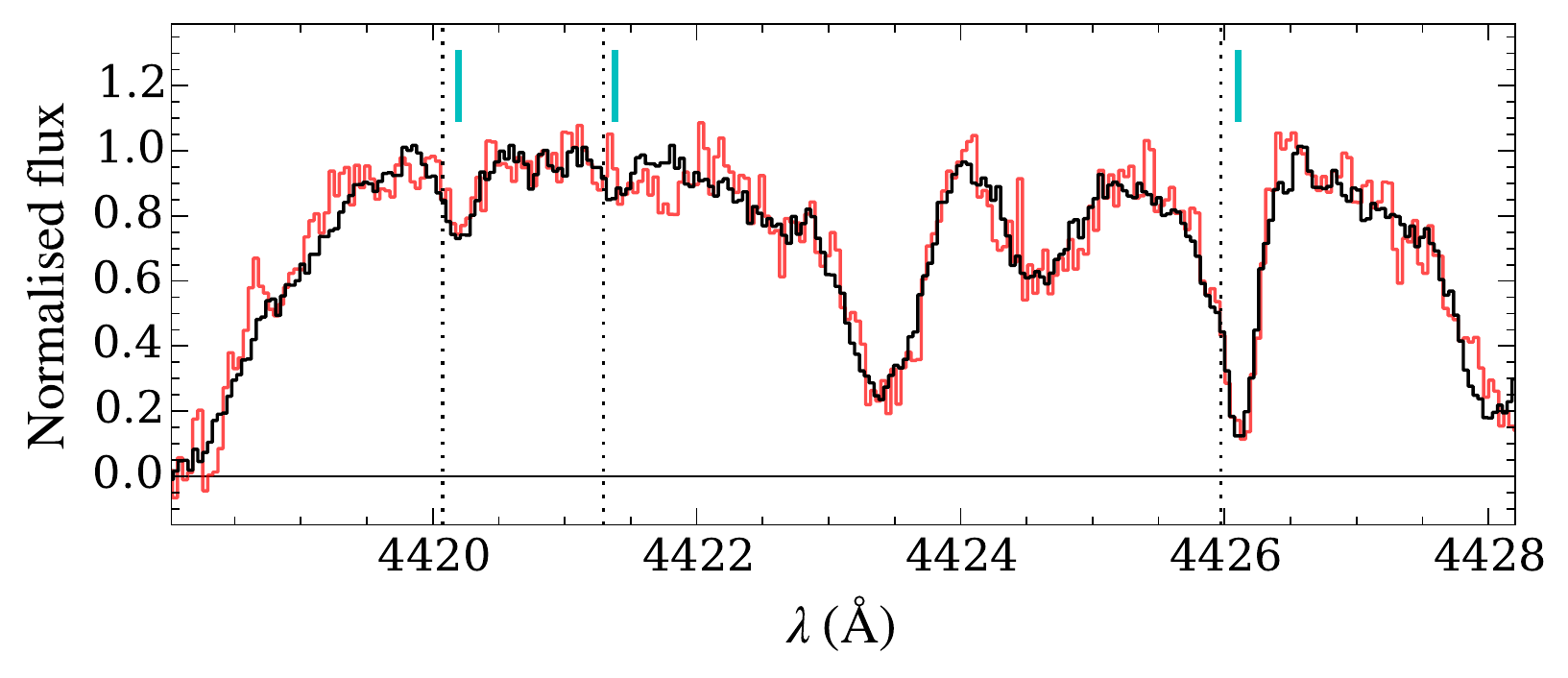}
\vspace{-0.3cm}
\caption{\label{f:shift}  The UVES (black) and HIRES (red) spectra
  of the \CIII\ 977 transition for all three components in
  LLS1249. Both spectra show a shift between the redshift of the
  \HI\ components, shown by dotted lines, and the metal line
  redshifts, shown by cyan ticks.}
\end{center}
\vspace{-0.3cm}
\end{figure}

\bsp	
\label{lastpage}
\end{document}